\documentstyle{article}

\def\be{\begin{equation}}
\def\ee{\end{equation}}
\begin{document}
\pagestyle{empty}

\vskip 4.0cm
\centerline{\bf{SMALL-\mbox{\boldmath $x$} PARTON DENSITIES FROM HERA
AND}}
\centerline{\bf{THE ULTRA-HIGH ENERGY
NEUTRINO-NUCLEON CROSS SECTIONS}}
\vskip 0.2true in
\centerline{Raj Gandhi$^{(1)}$, Chris Quigg$^{(2)}$, M. H. Reno$^{(3)}$
and Ina
Sarcevic$^{(4)}$\footnote{Talk presented by I. Sarcevic.}}
\vskip 0.15true in
\centerline{(1) {\em Mehta Research Institute,
10, Kasturba Gandhi Marg, Allahabad 211002, India}}
\centerline{(2) {\em Fermi National
Accelerator Laboratory, Batavia, IL 60510 USA}}
\centerline{(3) {\em Department of Physics and Astronomy, University of Iowa,
Iowa City, IA 52242 USA}}
\centerline{(4) {\em Department of Physics, University of Arizona,
Tucson, AZ 85721 USA}}

\vskip 2.0true in
\begin{abstract}
In light of recent measurements of the nucleon structure
function in the small-$x$ deep-inelastic regime at HERA
and the consequently
 improved theoretical understanding of the quark distributions in this
range of parton fractional momentum, we present new
results for the neutrino-nucleon cross
section at ultra high energies, up to $ E_{\nu} \sim 10^{21}$ eV.
The results are relevant
to deep underground muon detectors looking for such neutrinos.
For $\simeq 10^{20}$ eV
neutrinos, our cross sections are about a factor of $4$ to
$10$ times larger
than the previously reported results by
Reno, Quigg and Walker.  We discuss implications of this
new neutrino-nucleon cross section for
a variety of current and future neutrino detectors.
\end{abstract}
\vskip 0.15true in
\newpage
\baselineskip=16.5pt
As the number of extant and planned experimental facilities for
extra-terrestrial neutrino detection testifies, their importance in extending
the particle physics frontier beyond the standard model
is widely recognized \cite{Detect}.
At the highest energies, neutrinos are  decay products of pions produced
in cosmic accelerators and thus provide a direct window to the most
energetic processes in the universe. Neutrino astronomy also has important
advantages over  gamma-ray astronomy. In the energy range $10^{12}$ to $10^{20}
$ eV, the expected  air-cascade signature of the gamma-ray flux is dwarfed
by the cosmic-ray background. In the same energy range the flux of neutrino
by-products of the cosmic-ray interactions is very small, and the expected
neutrino flux from cosmic accelerators should dominate. In addition, neutrino
telescopes span a significant fraction of the sky at all times, in contrast
to gamma-ray detectors which provide small angular coverage for a limited
amount of time.

Recent observations \cite{AGN}
of TeV gamma-rays from Mkn 421 and Mkn 501 have revived
interest in studying mechanisms for producing high energy photons in
Active Galactic Nuclei (AGN). If the observed photons are
decay products of $\pi^0$s produced in hadronic interactions in
the disk surrounding the AGN, then AGNs are also powerful sources of
ultra high energy (UHE) neutrinos \cite{Ghs}.

In general, fluxes of UHE neutrinos fall
steeply with increasing energy,
making the direct detection of these particles difficult. Cerenkov
detectors, however, are capable of detecting upward-moving muons produced
by the charged-current (CC) interactions of energetic
neutrinos in the rock below the detector.
The effective volume of the detector is thus enhanced
in proportion to the range of the
produced muon (typically several kilometers), which then may traverse the
fiducial volume of the detector or stop therein.
The expected event rate for these detectors is presently subject to
two sources of uncertainty: $a)$ the uncertainty in the knowledge
of the incident neutrino flux, and, $b)$ the calculation of the
UHE charged
current cross-section for
$\nu_{\mu} + N \rightarrow \mu + X$, where $N$ is a
nucleon.
The latter stems from necessary extrapolations of the nucleon quark
structure functions for very low parton fractional momentum $x$
and large  momentum transfer $Q^2$.

Using recent results from the $ep$
collider at HERA \cite{He}
on small-$x$ parton distributions,
we have performed
calculation of the neutrino-nucleon cross sections for
neutrino energies up to $10^{21}$ eV \cite{Gqrs}.
For $10^{20}$ eV neutrinos
we have found
the cross sections to be
factor of four to ten times larger than previous estimates.  Since
UHE neutrinos are detected via neutrino-induced muons, our results
translate into significantly larger
probability
that neutrino-induced muon created in the rock surrounding
the detector would reach the detector and thus larger downward
muon rates.  For upward-moving muons,
the neutrino attenuation in the Earth depends on the
charged-current and
neutral-current
cross sections.  For larger neutrino-nucleon cross sections, the
attenuation effect is stronger.  Since it is the product of the muon
probability and the neutrino attenuation that goes into the
calculation of the upward-moving muon event rate, one has to
incorporate these
two competing effects, as elaborated below.

The inclusive cross section for $\nu_{\mu} + N \rightarrow
\mu^{-} + X$ is given by
\begin{equation}
\frac {d^2\sigma}{dxdy}=\frac{2G_F^2ME_\nu}{\pi}
\frac{M_W^4}{{(Q^2+M_W^2)}^2} (xq(x,Q^2)+x(1-y)^2\bar q(x,Q^2)),
\end{equation}
where $x=Q^2/2M\nu$, $y=\nu/E_\nu$,
and $\nu$  the energy loss in the lab frame, $\nu=E_\nu -E_\mu$.
The mass of the nucleon is $M$, and $M_W$ is the mass of the $W$-boson, while
the Fermi constant, $G_F=1.16 \times 10^{-5}$ GeV$^{-2}$.
With the assumption that the target is an isoscalar nucleon,
the
quark and antiquark
distribution functions,
$q(x,Q^2)$ and $\bar{q}(x,Q^2)$, contain valence and sea quark contributions.

At low energies, the neutrino-nucleon cross section reduces to the
usual 4-Fermi approximation and increases linearly with the neutrino energy.
At high energies, $E_\nu\gg M_W^2/2M$,
the $W$-propagator in Eq. (1) becomes important \cite{Rq}, limiting
the growth of $Q$ to $\langle Q^2\rangle \sim M_W^2$.
For $E_\nu\stackrel{>}{\sim}
10^5$ GeV,
the neutrino-nucleon total cross section is dominated by
the behavior of the quark densities at
$x\sim M_W^2/2ME_\nu$ and $Q^2\sim M_W^2$.
For  neutrino energies above
$10^{5}$ GeV, the
small-$x$ (i.e., $x\leq 3 \times 10^{-2}$)
behavior of the quark densities
plays an  important role in determining the cross-section.

Recently,
knowledge of the parton
densities at small values of $x$ has been substantially improved by the
observation of a
dramatic increase of the proton structure function,
$F_2(x,Q^2)$ with decreasing $x$ for $10^{-4}\leq x\leq 10^{-2}$
and $8.5$ GeV$^2 \leq Q^2 \leq 15$ GeV$^2$
by the
ZEUS and H1 Collaborations at HERA \cite{He}.
For larger values of $Q^2$,
the HERA experiments have measured $F_2^{ep}(x,Q^2)$
down to a value of $x= 2\times 10^{-2}$.
Thus,
for neutrino energies above $E_\nu\sim 10^5$ GeV,
the relevant small-$x$
regime is unconstrained by experiments.

In the absence of measurements
in the appropriate range of $(x,Q)$,
the parton distributions
are traditionally obtained by assuming certain form at
some fixed value of $Q=Q_0\approx$ few GeV
and the distributions are evolved to
larger values of $Q$ via
QCD evolution equations.
The parameters in the initial parametrizations
are determined by
fitting a
variety of experimental data on
structure functions for
$x>10^{-3}$ and moderate values of $Q$, as well as the recent
HERA data \cite{He}.

The theoretical
uncertainties in the total cross section for UHE neutrinos in the
standard model are therefore due to the parameterization at $Q_0$,
the extrapolation to values of $x$ below current data
and
in the evolution of the parton distribution functions to $Q>Q_0$.

Currently there are two
theoretical approaches
to  understanding small-$x$ evolution with $Q$,
both based on perturbative QCD.  The
first, and more traditional approach is to determine
parton densities for $Q>Q_0$ by solving the
next-to-leading order
Gribov-Lipatov-Altarelli-Parisi (GLAP)
evolution equations numerically \cite{Gl}.
Crucial to the calculation of the UHE $\nu N$ cross section is the
small-$x$ parameterization, which in all cases has
the form for sea quark distributions of $x\bar{q}(x,Q_0^2)\sim x^{-\lambda}$.
The second approach to small-$x$ evolution is to solve the
Balitskii-Fadin-Kuraev-Lipatov
(BFKL) equation \cite{Kl}. The solution predicts $\lambda \simeq 0.5$
and a more rapid evolution with $Q^2$ than the GLAP evolution.

The
parton distribution sets obtained
perturbatively using
next-to-leading order (NLO)
GLAP evolution equations are applicable for the calculation of
the total neutrino cross section for $E_\nu<10^5$ GeV, and are
a reasonable starting point for calculations for
higher energy neutrinos.
The NLO
CTEQ3 distributions \cite{CTEQ}
in the deep-inelastic scattering factorization
scheme (CTEQ-DIS) have
$Q_0=1.6$ GeV and
$\lambda = 0.332$.
These distributions are particularly useful because the
numerical evolution is provided for $x\rightarrow 0$, albeit, in a
region where GLAP evolution is unreliable.

\begin{figure}[htbp]
\vspace*{6.0 cm}
\caption{The charged current cross section for the CTEQ-DIS,
CTEQ-DLA, D\_ and EHLQ-DLA parton distribution functions.  The data point,
an average of ZEUS and H1,
is
from Ref. [12].}
\end{figure}

To estimate the uncertainty in the cross section, we consider
two other distribution functions extrapolated to small-$x$. An approximate
lower limit is given by the leading order CTEQ3 distribution
with the double-logarithmic-approximation (DLA)
extrapolation below $x_{\rm min}=10^{-4}$,
labeled CTEQ-DLA below. This is an approximate solution to
the leading order GLAP equations, suitable for $\lambda\sim 0$.
We use a more
singular distribution than the CTEQ-DIS distributions
to indicate the upper range of the cross section.
Motivated by BFKL dynamics,
the MRS D\_ set \cite{Ms}
is parameterized with $\lambda = 0.5$.
The power $\lambda=0.5$ does not change
significantly with $x$ and $Q$ for small-$x$ and $Q \sim M_W$, so
we use a power law extrapolation below the limit of the
numerically evolved distributions, namely for
$x<x_{\rm min}=10^{-5}$.

In Fig. 1 we present our results for the
charged-current cross sections
for the three
choices of parton distribution functions, and compare with
the calculation of Ref. \cite{Rq} which used the Eichten
{\it et al.} \cite{Ehlq} distributions extrapolated using the
DLA (EHLQ-DLA).
We find our results
to be in very good agreement
with the recent measurement of charged-current cross section at
HERA \cite{Cc}.

The differences in the cross section appear at
neutrino energies $E_\nu\sim 10^6$ GeV, reflecting the different
behaviors of $xq_s(x,Q)$ below $x\sim M_W^2/(2 M \cdot 10^6$ GeV)$\sim
3\cdot 10^{-3}$. The D\_ and CTEQ-DIS cross sections
are essentially equal for $E_\nu=10^6$ GeV, but the CTEQ-DLA cross section
is $\sim 15\%$ lower. By $E_\nu=10^{10}$ GeV, there is a factor of 2.6
difference between the CTEQ-DLA extrapolation and the D\_ calculations of the
cross section. The CTEQ-DIS result lies a factor of 1.4 above the CTEQ-DLA
cross section. At $E_\nu=10^{12}$ GeV, the discrepancy between the
CTEQ-DLA and D\_ calculations is a factor of 6.
One should keep in mind, however, that the DLA is an
approximation valid for $E_\nu \ll 10^{11}$ GeV.

In order to calculate the number of upward-moving muons that can
be detected with future neutrino detectors, such as DUMAND II,
AMANDA,
BAIKAL and NESTOR \cite{Detect}, we fold in the muon flux and its attenuation
due to its passage through the Earth with the probability that a neutrino
passing on a detector trajectory
creates a muon in the rock which traverses the detector.
The cross section appears in the exponent for the attenuation, and
the probability is proportional to the cross section. The
details are found in Ref. \cite{Gqrs}.
We find that above $10^5$ GeV, both the muon probability and the
neutrino attenuation are very sensitive to the small-$x$ behavior
of the sea quark distributions, but their product is not.
Thus, for the downward-moving muons (and for the contained events) the
event rates obtained with MRS D\_
or CTEQ distributions are much larger
than the same with EHLQ structure functions, but the upward rates
differ only by $17\%$ for the whole energy range.

We find that future detectors, such as
DUMAND II, AMANDA, BAIKAL
and NESTOR have a very good chance of being able to test different models for
neutrino production in the AGNs \cite{Ghs}.
For $E_{\mu}>1$ TeV, we find that
900
upward-moving
muons/yr/km$^2$/sr
originating from
the diffuse AGN neutrinos, with the atmospheric background of 1400
events/yr/km$^2$/sr. 
For $E_{\mu}>10$ TeV, signal to background ratio becomes even better,
the signal of $500$ muon events/yr/km$^2$/sr,
being almost 20 times the background rate.


\begin{thebibliography}{99}
\baselineskip=13pt
\bibitem{Detect}J. Babson {\it et al.}, (DUMAND Collaboration), {\it Phys.
Rev. } {\bf D42} (1990) 3613;
{\it Proceedings of the NESTOR workshop at Pylos, Greece},
ed. L.K. Resvanis (University of Athens, 1993);
D. Lowder {\it et al.}, {\it Nature} {\bf 353} (1991) 331.
\bibitem{AGN} M. Punch, {\it et al.} (Whipple Observatory Gamma Ray
Collaboration) {\it Nature (London)} {\bf 160} (1992) 477;
A.D. Kerrick {\it et al.}, {\it Astrophys. J.} {\bf 438} (1995) L59;
J. Quinn, {\it et al.} (Whipple Observatory Gamma Ray Collaboration)
{\it IAU Circular} 6169 (June 16, 1995).
\bibitem{Ghs} See T. Gaisser, F. Halzen and T. Stanev,
{\it Physics Reports} {\bf 258} (1995) 173
for a review of neutrino astronomy and
sources of UHE neutrinos. For the event rates presented here, we
use the AGN model of
F. Stecker {\it et al., Phys. Rev. Lett.} {\bf 66} (1991) 2697;
{\it ibid.} {\bf 69} (1992) 2738(E) as one example.  The background
atmospheric flux is taken from L.V. Volkova, {\it Sov. J. Nucl. Phys.}
{\bf 31} (1980) 784.
\bibitem{He}  ZEUS Collaboration, M. Derick {\it et al.}, {\it Phys. Lett.}
{\bf B316} (1993) 412; H1 Collaboration, I. Abt {\it et al.},
{\it Nucl. Phys.} {\bf B407} (1993) 515.
\bibitem{Gqrs}  R. Gandhi, C. Quigg, M.H. Reno and I. Sarcevic,
FERMILAB-PUB-95/221-T.
\bibitem{Rq}  C. Quigg, M.H. Reno
and T. Walker, {\it Phys. Rev. Lett.} {\bf 57}
(1986) 774; M.H. Reno and C. Quigg, {\it Phys. Rev.} {\bf D37} (1988)
657.
\bibitem{Gl}
V.N.~Gribov and L.N.~Lipatov, {\it Sov. J. Nucl. Phys.} {\bf 15} (1972)
438; L.N.~Lipatov, {\it Sov. J. Nucl. Phys.} {\bf 20} (1974) 181;
Yu.L.~Dokshitser, {\it Sov. Phys. JETP} {\bf 46} (1977) 641;
G.~Altarelli and G.~Parisi, {\it Nucl. Phys.} {\bf B126} (1977) 298.
\bibitem{Kl}
E.A.~Kuraev, L.N.~Lipatov and V.S.~Fadin, {\it Sov. Phys. JETP}
{\bf 44} (1976) 443; {\bf 45} (1977) 199;
Ya.Ya.~Balitskii and L.N.~Lipatov, {\it Sov. J. Nucl. Phys.} {\bf 28} (1978)
822.
\bibitem{CTEQ}
H. Lai {\it et al.}, {\it Phys. Rev.} {\bf D51} (1995) 4763.
\bibitem{Ms} A.D. Martin, W.J. Stirling and
R.G. Roberts, {\it Phys. Rev.} {\bf D47} (1993) 867.
\bibitem{Ehlq}  E. Eichten, I. Hinchliffe, K. Lane and
C. Quigg, {\it Rev. Mod. Phys.} {\bf 56} (1984) 579.
\bibitem{Cc}
Rolf Beyer, 1995 {\it Workshop on Weak Interactions and Neutrinos},
Talloires,
France.


\end{thebibliography}
\end{document}